\documentclass[twocolumn,showpacs,aps,prl]{revtex4}  
\usepackage{epsfig,amssymb}

\begin{document}

\title{Formation of Quantum Shock Waves by Merging and Splitting Bose-Einstein Condensates}

\author{J.J. \surname{Chang}}
\author{P. \surname{Engels}}
\email{engels@wsu.edu} \affiliation{Washington State University, Department of Physics and Astronomy, Pullman,
Washington 99164, USA}
\author{M. A. \surname{Hoefer}}
\affiliation{National Institute of Standards and Technology, Boulder, Colorado 80305}
\thanks{This contribution of NIST, an agency of the U.S. government, is not subject to copyright.}

\date{\today}

\begin{abstract}

The processes of merging and splitting dilute-gas Bose-Einstein condensates are studied in the nonadiabatic,
high-density regime. Rich dynamics are found. Depending on the experimental parameters, uniform soliton trains
containing more than ten solitons or the formation of a high-density bulge as well as quantum (or dispersive)
shock waves are observed experimentally within merged BECs. Our numerical simulations indicate the formation of
many vortex rings. In the case of splitting a BEC, the transition from sound-wave formation to dispersive
shock-wave formation is studied by use of increasingly stronger splitting barriers. These experiments realize
prototypical dispersive shock situations.

\end{abstract}

\pacs{03.75.Kk,67.85.De,05.45.-a,47.40.-x} \maketitle

\par
Dilute-gas Bose-Einstein condensates (BECs) are a powerful environment for the study of linear and nonlinear
dynamics. In this context, the merging and splitting of BECs are two particularly important processes that,
despite their apparent simplicity, provide access to a variety of different regimes. For example, the
interference fringes observed when two BECs collide in the low-density regime reveal the fundamental matter wave
nature underlying BECs \cite{matterwaves,Reinhardt1997}. In contrast, numerical studies of fast collisions in
the high-density regime reveal indications of quantum turbulent behavior \cite{Norrie2005}. Slower merging in
the high-density regime has experimentally been shown to lead to vortex formation \cite{Anderson2007,
Cornell2007}, providing information about the phase evolution of the merging BECs. Theoretical studies
demonstrate that adiabatic merging of BECs can lead to soliton formation, too \cite{Kivshar2007}. A precise
understanding of the merging dynamics is also essential from a technological point of view. For example,
splitting and merging processes are fundamental operations in atom interferometers \cite{atominterferometry}.
The merging of BECs is also an important process in the creation of a ``continuous BEC'' \cite{Ketterle2002}
where a condensate was continuously replenished by newly condensed atoms.

In this Letter, we study the physics of merging and splitting BECs in the nonadiabatic, high-density regime and
use these processes to investigate dispersive hydrodynamics, including dispersive shock waves (DSWs, also
referred to as quantum shock waves). After merging two BECs, we can observe many solitons (a soliton train). For
low enough atom numbers, the soliton train is uniform, as predicted for a one-dimensional situation
\cite{Reinhardt1997,Brazhnyi2003,Damski2006}. For higher atom numbers, a high-density bulge emerges, and our
numerical simulations suggest that this bulge consists of many vortex rings due to a transverse instability of
the soliton train. Splitting a BEC with a repulsive barrier can lead to DSWs, and we observe a transition from
propagating sound waves to DSWs when a sufficiently strong barrier is used. Finally, we also find shock dynamics
in a different setting, namely when a high-density region in a BEC is suddenly released and allowed to spread
into a surrounding background of condensed atoms. In this case, the initial high-density peak splits into two
peaks that propagate away from each other, and after a sufficient self-steepening of the wavefront, DSWs are
formed. These are important prototypical situations that have been discussed in previous theoretical studies
\cite{Damski2006,Hoefer2007,Damski2004,BECshocktheory,Hoefer2006}, and our results complement previous
experiments that considered either very narrow initial gaps in a BEC, produced by a stopped-light technique
\cite{Hau2001,Ginsberg2005}, or blast pulses in rotating \cite{Simula2005} and nonrotating \cite{Hoefer2006}
cylindrical geometries. In addition, we demonstrate intriguing similarities between the hydrodynamics of BECs
and the propagation of laser light through nonlinear crystals, where related dispersive shock phenomena have
also been observed \cite{Fleischer2007}.

All our experiments begin with ultracold clouds of $^{87}$Rb atoms in the $|F,m_{F}\rangle$ = $|1,-1\rangle$
hyperfine state. The atoms are magnetically contained in an elongated Ioffe-Pritchard type trap with frequencies
$\{\omega_{x}/(2\pi),\omega_{yz}/(2\pi)\}=\{7, 402\}$~Hz (the $x$-axis is oriented horizontally). Repulsive and
attractive barriers for the atoms are created with dipole lasers that are far detuned from the Rb D-lines at
780~nm and 795~nm. The dipole laser beams are sent horizontally through the center of the magnetic trap, along
the radial (tightly confining) $y$-direction. In the vertical direction ($z$-axis), the laser waist is much
larger than the radial extent of the BECs. Dynamics are induced in the BEC by rapidly turning a dipole beam on
or off. To enlarge the resulting features we employ an antitrapped expansion 2~ms long before imaging
\cite{Lewandowski2003}. During this expansion, the aspect ratio of a BEC formed without the presence of a dipole
barrier changes from 57 for the trapped BEC to an aspect ratio of about 3.

\begin{figure} \leavevmode \epsfxsize=3.375in
\epsffile{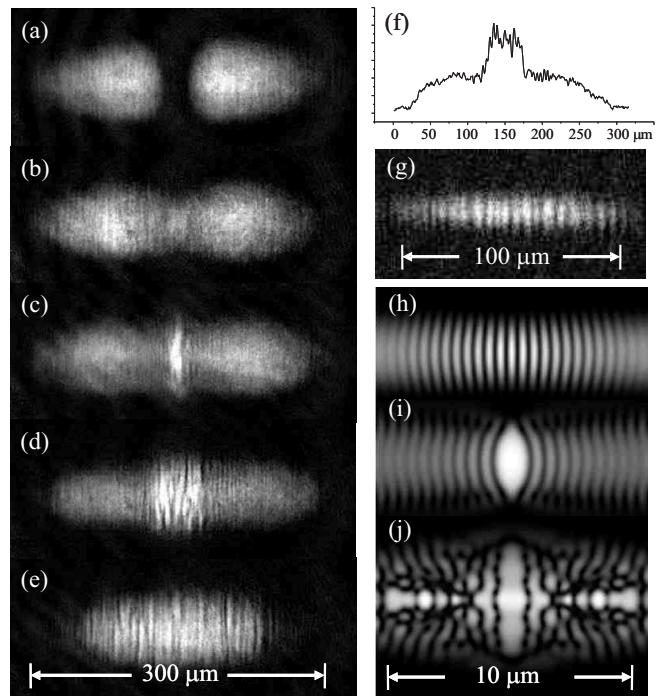} \caption{\label{singlepokyseq} Left column: experimental antitrapped expansion images
of a BEC collision at $t=$ (a)~0~ms, (b)~5~ms, (c)~10~ms, (d)~20~ms, and (e)~55~ms. Right column: (f) integrated
cross sections of (d). (g) Typical uniform soliton train observed for lower atom numbers (experimental image;
parameters see text). (h-j) Numerical simulations with the same initial parameters as in left column. Zoomed-in
views after (h)~5~ms, (i)~6.25~ms, and (j)~7.5~ms. The simulations show density slices of the BEC by the plane
$z = 0$. No antitrapped expansion was performed in the simulations, see also \cite{EPAPS}. }
\end{figure}

In a first set of experiments, the dynamics of merging two BECs in the nonadiabatic, high-density regime are
studied. For this, a dipole beam with a wavelength of $\lambda=660$~nm, a power of 3.48~mW and waists of
$w_{x}=27.3~\mu$m and $w_{z}=32.1~\mu$m is used, creating a repulsive barrier with a height of 490~nK for the
atoms. The total atom number for the experiments shown in Figs.~\ref{singlepokyseq}(a-f) is about $1 \times
10^6$ atoms. For a single BEC confined in the magnetic trap, this would imply a chemical potential of
$\mu=224$~nK. Therefore the presence of the dipole beam leads to two clearly separated BECs
(Fig.~\ref{singlepokyseq}(a)). The beam is turned on before the atoms are evaporatively cooled to form a BEC.
After a BEC has formed on both sides of the barrier and no surrounding thermal cloud is visible, the dipole beam
is rapidly turned off within less than 250~ns. We let the dynamics evolve in the magnetic trap for a variable
evolution period before starting the expansion imaging. Directly after turning the dipole barrier off, the
condensates smoothly expand toward each other (Fig.~\ref{singlepokyseq}(b)).  This behavior can be described by
the well-known dam-breaking problem whereby a sharp density gradient develops into a rarefaction wave (as
opposed to a shock wave) when the background density is zero (see e.g. \cite{Hoefer2006}). Shortly after the
BECs have collided at the center of the trap, a pronounced bulge of higher atom density forms in the collision
plane (Fig.~\ref{singlepokyseq}(c)). Subsequently this density bulge spreads out from the center of the trap
(Figs.~\ref{singlepokyseq}(c-e)). Very pronounced dark notches are observed to form within the high-density
bulge as shown in Fig.~\ref{singlepokyseq}(d). As the bulge spreads out from the collision plane, more and more
notches are formed to fill the extent of the density bulge with an average spacing of roughly 8~$\mu$m to
11~$\mu$m. After about 55~ms, the bulge and the notches have spread over the entire extent of the condensate
(Fig.~\ref{singlepokyseq}(e)). The long lifetime, discrete nature, and large amplitude of the notches suggests
that they are nonlinear coherent structures rather than simple sound waves. Our numerical simulations show that
a soliton train initially develops and a bulge region is formed where the solitons decay into a large number of
vortex rings, see Figs.~\ref{singlepokyseq}(h-j) and \cite{EPAPS}. Experimentally, vortex rings in BECs have
been observed in \cite{Anderson2001,Ginsberg2005}. They are difficult to detect unambiguously in our
experimental images that are integrated along the line of sight. Fine fringes appear adjacent to the bulge
region as can be seen, e.g., in Fig.~\ref{singlepokyseq}(d). The fine fringes, together with the steepness of
the wavefronts delimiting the density bulge region, are indicative of DSWs. The merging process finally results
in an axial breathing-mode excitation of the BEC.

\begin{figure} \leavevmode \epsfxsize=3.375in
\epsffile{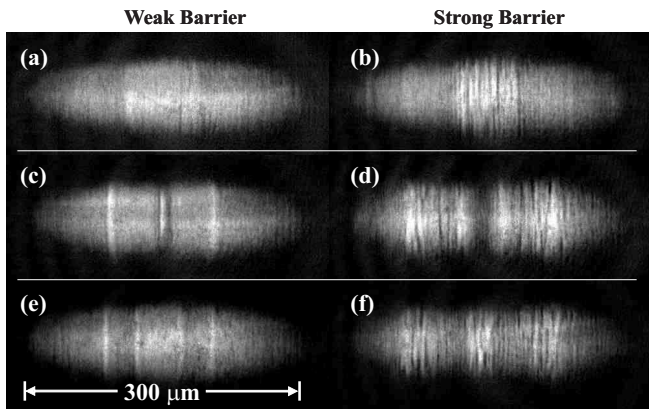} \caption{\label{picturematrix} Dynamics induced by repulsive barrier. (a), (b)
Turning a barrier off after forming a BEC in the presence of the barrier. (c), (d) Turning a barrier on after
forming a BEC without the presence of the barrier. (e), (f) Pulsing the barrier on for 1.5~ms after the
formation of a BEC without the presence of the barrier. (a), (c), (e) Weak barrier, laser power 360~$\mu$W. (b),
(d), (f) Strong barrier, laser power 1.99~mW. Evolution times before start of imaging procedure: (a),(b)~16~ms,
(c), (d)~12~ms, (e), (f)~10~ms.}
\end{figure}

The qualitative features of the evolution are fairly independent of most experimental parameters. For example,
use of two BECs with an initial total atom number of $2.2 \times 10^6$ atoms and a dipole beam with waists of
$w_{x}= 8.5 ~\mu$m and $w_{z}=32.1~\mu$m gives qualitatively the same results (see Fig.~\ref{picturematrix}(b)).
However, when the atom number is strongly reduced, we observe a transition in the merging dynamics, both
experimentally and numerically, from the generation of a high-density bulge to the generation of a uniform
soliton train with no bulge. A typical image of such a  soliton train is shown in Fig.~\ref{singlepokyseq}(g)
for 22000 atoms, a dipole beam power of $150~\mu$W, an evolution time of 27~ms and an expansion time of 1~ms
(see also \cite{EPAPS}). A one-dimensional analysis akin to that in \cite{Hoefer2007} reveals that the soliton
train can be interpreted as the result of the interaction of two degenerate rarefaction waves resulting from the
dam-breaking problem. A detailed analysis of this transition is beyond the scope of this work.

\begin{figure} \leavevmode \epsfxsize=3.375in
\epsffile{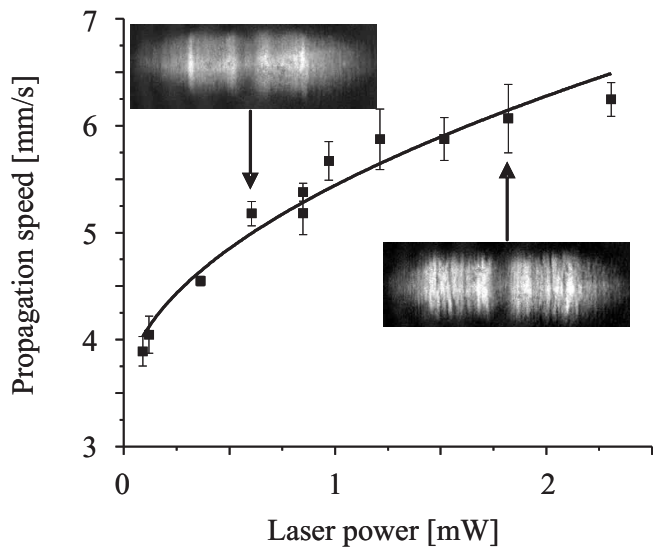} \caption{\label{soundspeed} Speed of wavefront propagation through a BEC. After a BEC
is formed, a repulsive dipole laser is suddenly turned on. The plot shows the propagation speed of the resulting
wavefronts vs. the applied laser power. Error bars are taken from fits of distances vs. time. The line shows the
function $v(P)=a+b \sqrt{P}$ fitted to the data. Insets show two representative images obtained with an
evolution time of 12.5~ms and a laser power respectively below and above the power at which the wavefronts start
breaking into solitons.}
\end{figure}

In a second set of experiments, we investigated the dynamics of splitting a BEC with a repulsive barrier that is
suddenly turned on in the center of the BEC. For very weak barriers that only slightly modify the BEC density,
the sudden turn-on leads to the propagation of sound waves \cite{Ketterle1997,Thomas2007}. Strong barriers, in
contrast, lead to DSWs. We first create BECs with $2.2\times10^6$ atoms in the magnetic trap without the
presence of the dipole barrier. Then a dipole beam with waists of $w_{x}= 8.5~\mu$m and $w_{z}=32.1~\mu$m is
rapidly turned on and left on for a variable evolution time, after which the antitrapped expansion procedure is
started. The rapid turn-on of the dipole beam produces two density peaks that spread out to either side, as
shown in Fig.~\ref{picturematrix}(c). The measured propagation speed of these peaks in the central region of the
BECs is plotted in Fig.~\ref{soundspeed} for various powers of the dipole beam \cite{footnotesoundspeed}. For
the lowest powers, the speed is in full agreement with the calculated longitudinal speed of sound, 3.8~mm/s
\cite{longsound}. At these low powers, the density peaks are barely visible in the cloud. For stronger dipole
beams, the propagation speed increases above the speed of sound, and for sufficiently strong beams the obtained
images change qualitatively due to dispersive shock formation with two associated speeds. For laser powers above
approximately 0.6~mW, fine fringes appear in front of the propagating peaks. When the dipole beam exceeds a
power of roughly 1.2~mW, solitons are formed in the region between the two wavefronts, as shown in
Fig.~\ref{picturematrix}(d). For an oblate, rather than cigar-shaped geometry, related ring-shaped structures
have been interpreted in the context of DSWs in \cite{Hoefer2006}. Numerical simulations \cite{EPAPS} suggest
that this behavior is qualitatively described as follows.  Two peaks in the density and outward superfluid
velocity are generated by the repulsive dipole beam. These peaks break when the ``entropy'' condition is
satisfied \cite{Damski2004}, causing the generation of two DSWs on the inner and outer edges of each peak. Due
to a transverse instability, these quasi-one-dimensional DSWs break up into many vortex rings, leading to
interactions. The interaction of DSWs has been studied in \cite{Hoefer2007}. Deriving an analytic expression for
the DSW speeds is complicated by the generation of vortex rings and wave interactions. Empirically, the
dependence of pulse propagation speed on the dipole laser power is described reasonably well by a square-root
dependence, as shown by the fitted curve in Fig.~\ref{soundspeed}.

Our experiments reveal that rapidly switching on a weak dipole barrier merely leads to two density peaks that
spread out (Fig.~\ref{picturematrix}(c)), whereas for strong barriers, soliton formation is observed in the wake
behind the wavefronts (Fig.~\ref{picturematrix}(d)). A similar transition also exists for the merging of BECs:
If two BECs are initially separated by a strong barrier, soliton dynamics are observed after turning the barrier
off, as described in detail above (Fig.~\ref{singlepokyseq} and Fig.~\ref{picturematrix}(b)). If, however, the
initial dipole barrier is so weak that it merely produces a small density suppression in an initial BEC, a
density bulge that does not contain solitons (Fig.~\ref{picturematrix}(a)) appears after the turn-off. Pulsing a
barrier on for a short time combines the effects of turning on and turning off a barrier
(Figs.~\ref{picturematrix}(e), (f)); see also \cite{EPAPS}.

\begin{figure} \leavevmode \epsfxsize=3.375in
\epsffile{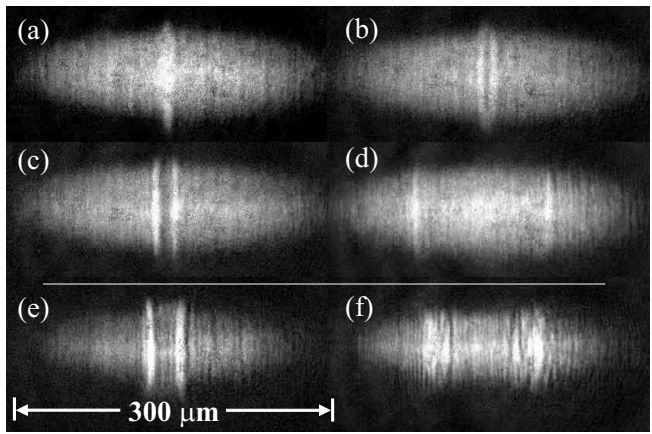} \caption{\label{dimple} Dynamics induced by suddenly turning off a local, attractive
dipole potential in the center of a BEC. Dipole laser wavelength 830~nm, power 61~$\mu$W. Evolution time between
turn-off and start of expansion: (a)~0~ms, (b)~0.25~ms, (c)~1~ms, (d)~14~ms. (e), (f) Development of dispersive
shock waves when a laser power of 183~$\mu$W is used. Evolution time (e) 1~ms, (f) 8~ms. }
\end{figure}

Yet another, different prototypical DSW situation comes about when a local high-density peak in a BEC is
suddenly released and is allowed to spread out into a surrounding background of lower density. We can realize
this experimentally by replacing the 660~nm dipole laser with an 830~nm laser, leading to an attractive dipole
potential. In Figs.~\ref{dimple}(a-d) such an attractive dipole beam with a waist of $w_{x}= 5~\mu$m and $w_{z}
= 41~\mu$m and a power of 61~$\mu$W is sent through the center of the magnetic trap in the radial direction.
Evaporative cooling in the presence of the combined optical and magnetic potential leads to a BEC with a
localized high-density peak in its center (Fig.~\ref{dimple}(a)). When the dipole beam is suddenly switched off
after the formation of the BEC, the central high-density peak spreads out into the surrounding parts of the BEC.
Interestingly, the high density peak does not remain a single peak but quickly splits into two peaks
(Fig.~\ref{dimple}(b)) that subsequently travel outward in opposite directions (Figs.~\ref{dimple}(c,~d)).
Indeed such splitting is expected from theory \cite{Damski2004} and has been observed in nonlinear optics
\cite{Fleischer2007}. Following \cite{Hoefer2007}, one can show that in the quasi-one-dimensional regime two
counterpropagating DSWs interacting with trailing rarefaction waves are generated in such a situation. For the
parameters in Figs.~\ref{dimple}(a-d), the measured propagation speed of the two propagating peaks is 4.34~mm/s.
When the initial dipole beam power is reduced, the measured speed decreases and approaches the longitudinal
speed of sound. When a stronger dipole beam with a power of 183~$\mu$W is used, again two peaks form and spread
out (Fig.~\ref{dimple}(e)), but now DSWs form, marked by solitons in the inner region and strong ripples in the
outer regions of the BEC (Fig.~\ref{dimple}(f)); see also \cite{EPAPS}.

In summary, our experiments realize several prototypical situations for dispersive shock wave formation and
demonstrate the transition from sound wave propagation to dispersive shock dynamics as increasingly stronger
perturbations are applied. Several aspects of this behavior, such as the splitting of an initial single peak
into two (Fig.~\ref{dimple}) followed by the formation of dispersive shock waves, are very similar to those
observed in nonlinear optics. This demonstrates the generality of our results and showcases the usefulness of
BECs in the study of nonlinear wave dynamics.

\begin{acknowledgments}
PE acknowledges financial support from NSF under Grant No. PHY-0652976.
\end{acknowledgments}

\onecolumngrid
\appendix*
\pagebreak
\section{Supporting material: numerical simulations}
\pagebreak

\begin{figure*}
  \centering
  \begin{tabular}{ccc}
    \begin{minipage}{0.315\linewidth}
      \includegraphics[width=\linewidth]{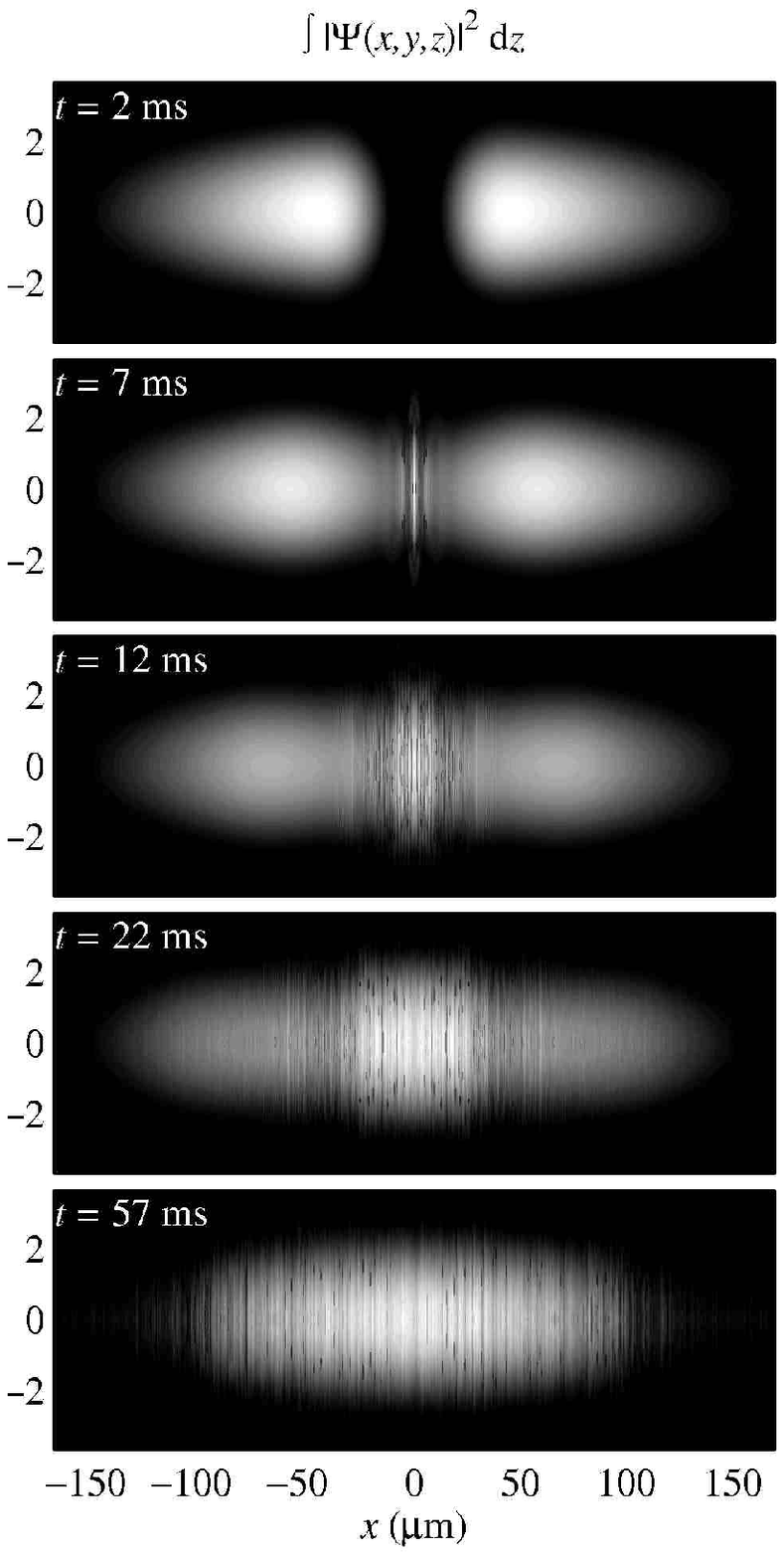}
    \center{(a)}
    \end{minipage}
    &
    \begin{minipage}{0.315\linewidth}
      \includegraphics[width=\linewidth]{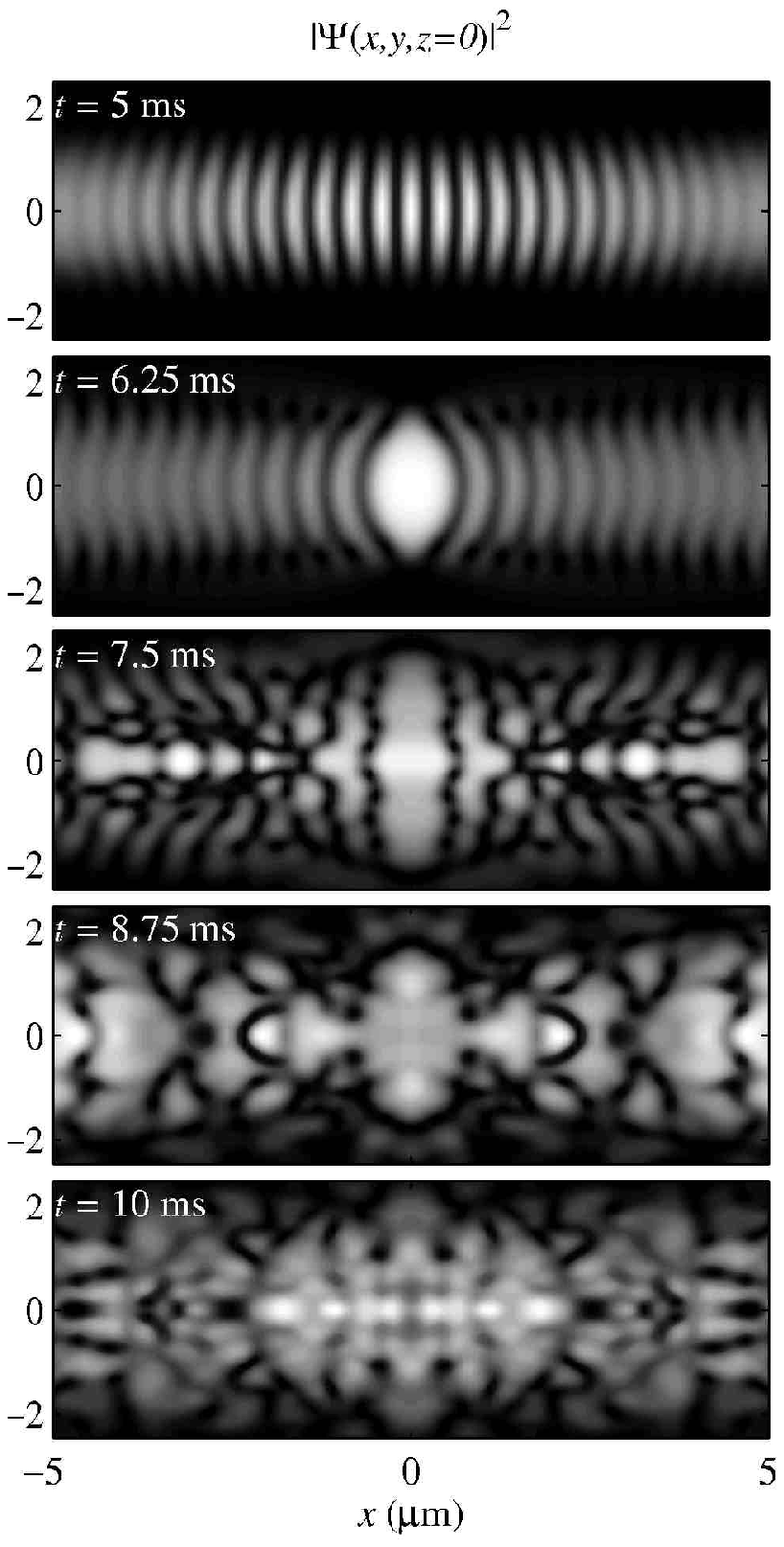}
    \center{(b)}
    \end{minipage}
    &
    \begin{minipage}{0.315\linewidth}
      \includegraphics[width=\linewidth]{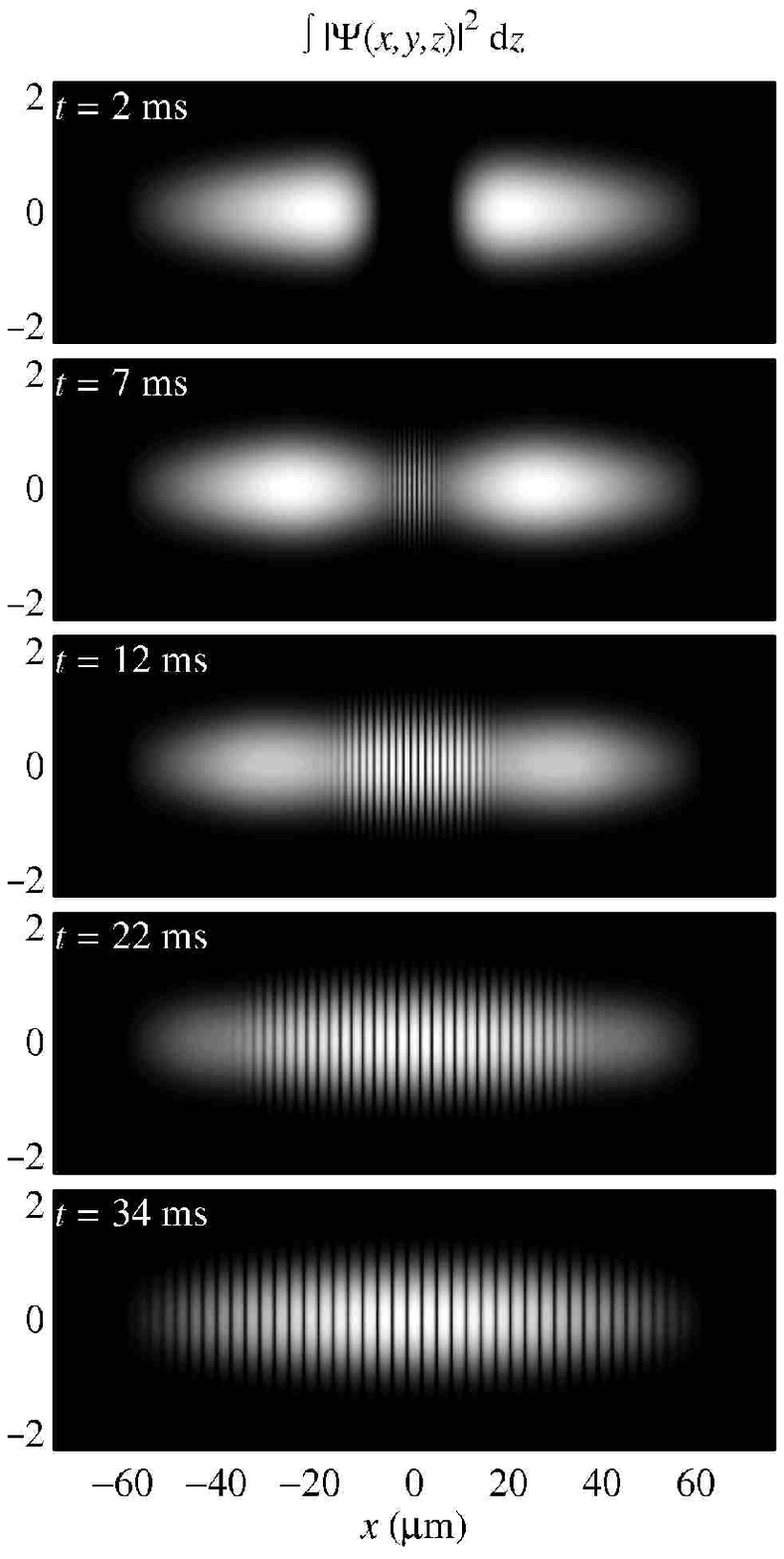}
      \center{(c)}
    \end{minipage}
  \end{tabular}
  \caption{Azimuthally symmetric numerical simulations of the
    Gross-Pitaevskii equation for the merging experiment.  Figs.\ (a)
    and (b) were generated using the same experimental parameters used
    to create Fig.\ 1 in the main text with $N = 10^6$ particles.
    Fig.\ (a) is the integrated density.  Fig.\ (b) represents density
    slices of the condensate by the plane $z = 0$.  This is a zoomed
    in version of the initial interaction process showing complicated
    spatial structure along with the generation of a large number of
    vortex rings which appears to be due to a transverse
    (snake) instability of the soliton train. The snake instability of
    a single dark soliton was studied, e.g., in D. L. Feder et al.,
    Phys. Rev. A 62, 053606 (2000). Fig.\ (c) shows density slices as
    in (b) and involves the same harmonic trapping parameters as in Figs.\ (a)
    and (b) but with fewer particles, $N = 18,600$ (approximately the same
    as for Fig.\ 1(g) in the main text).  Quasi-one-dimensional behavior
    is preserved and a soliton train is generated in this case.  No
    anti-trapped expansion was performed in either simulation.  The vertical,
    $y$-axis in the figures is measured in units of microns.}
\end{figure*}
\pagebreak
\begin{figure*}
  \centering
  \begin{tabular}{ccc}
    \begin{minipage}{0.315\linewidth}
      \includegraphics[width=\linewidth]{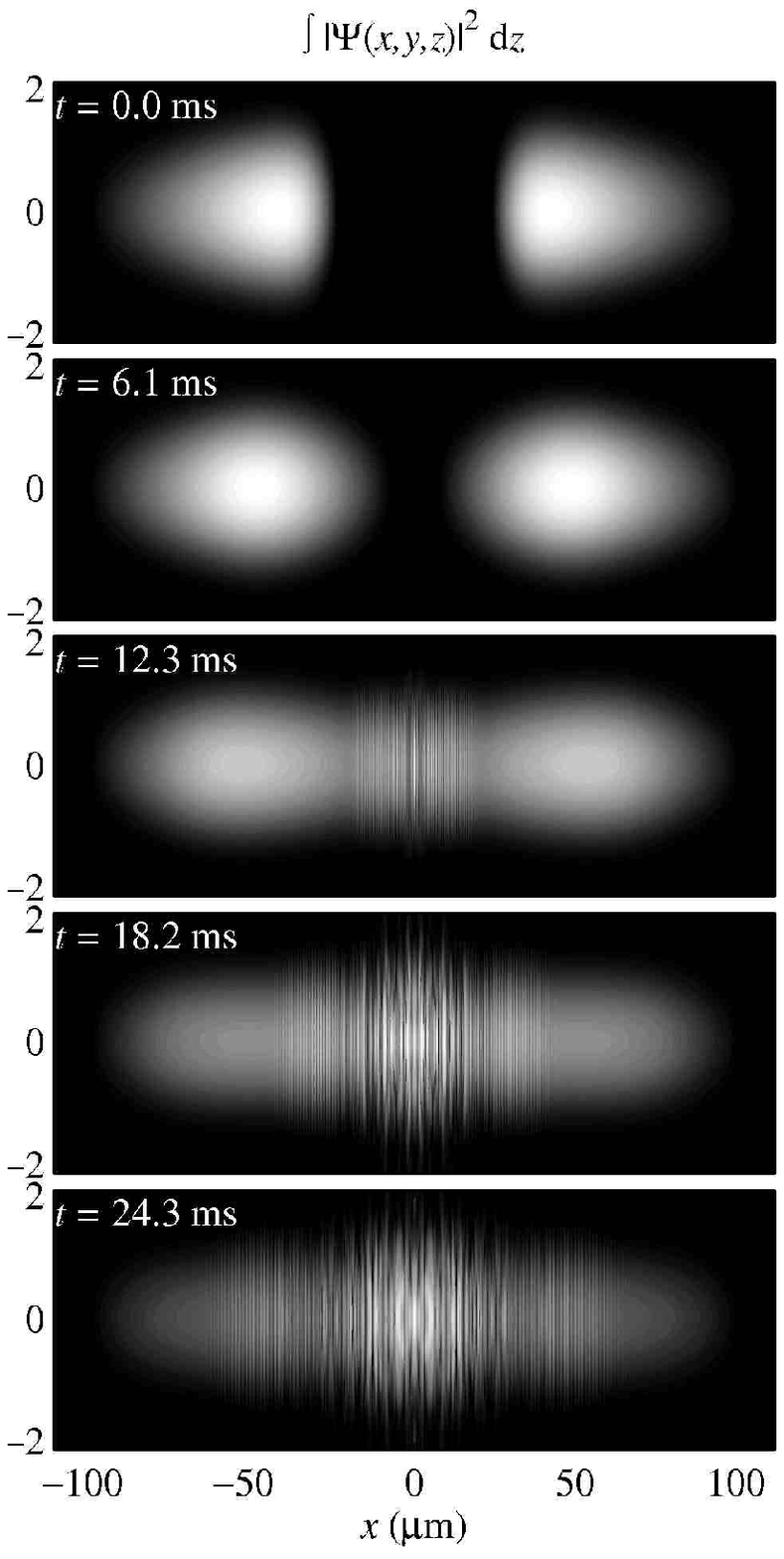}
      \center{(a)}
    \end{minipage}
    &
    \begin{minipage}{0.315\linewidth}
      \includegraphics[width=\linewidth]{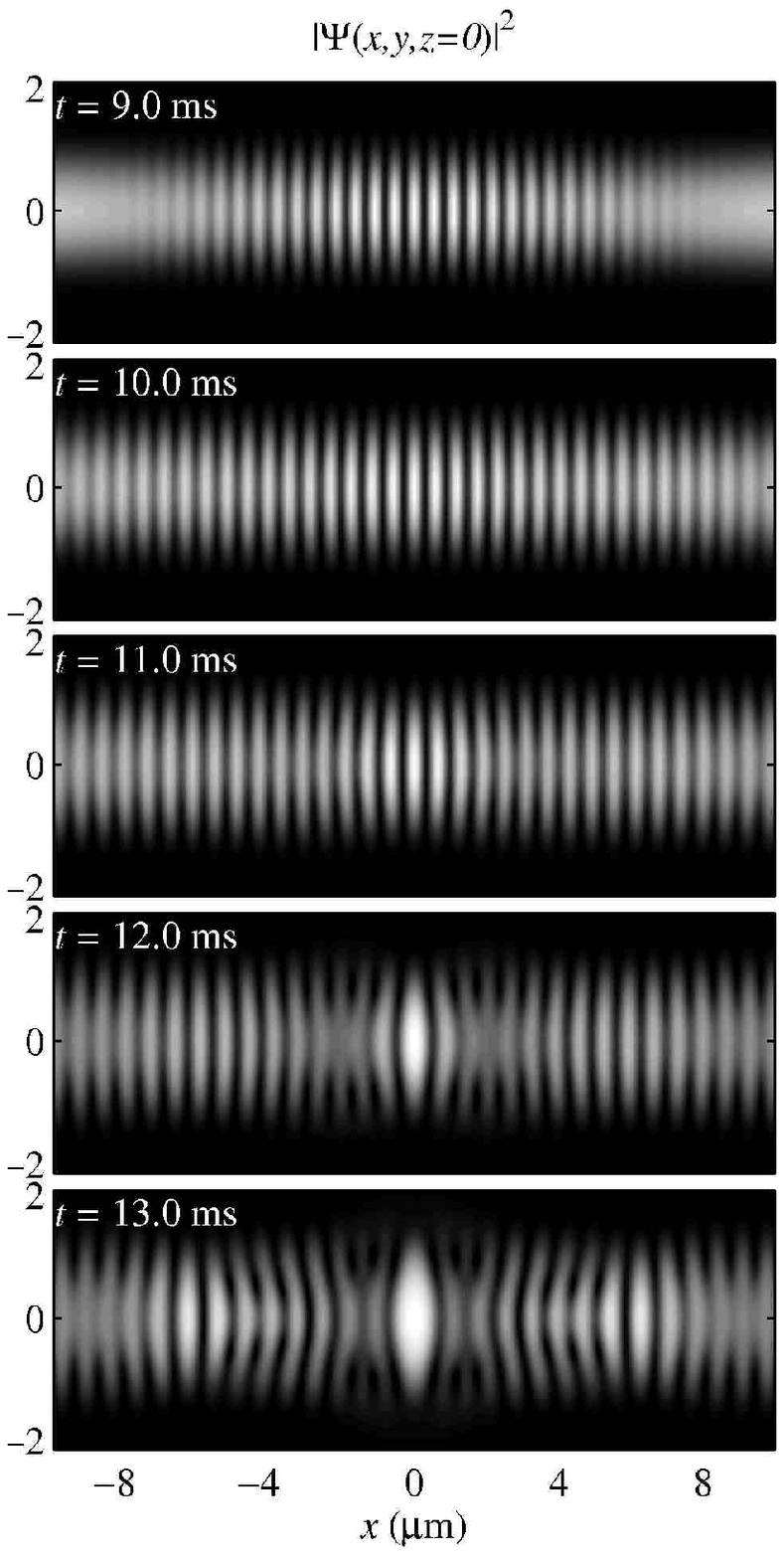}
      \center{(b)}
    \end{minipage}
    &
    \begin{minipage}{0.315\linewidth}
      \includegraphics[width=\linewidth]{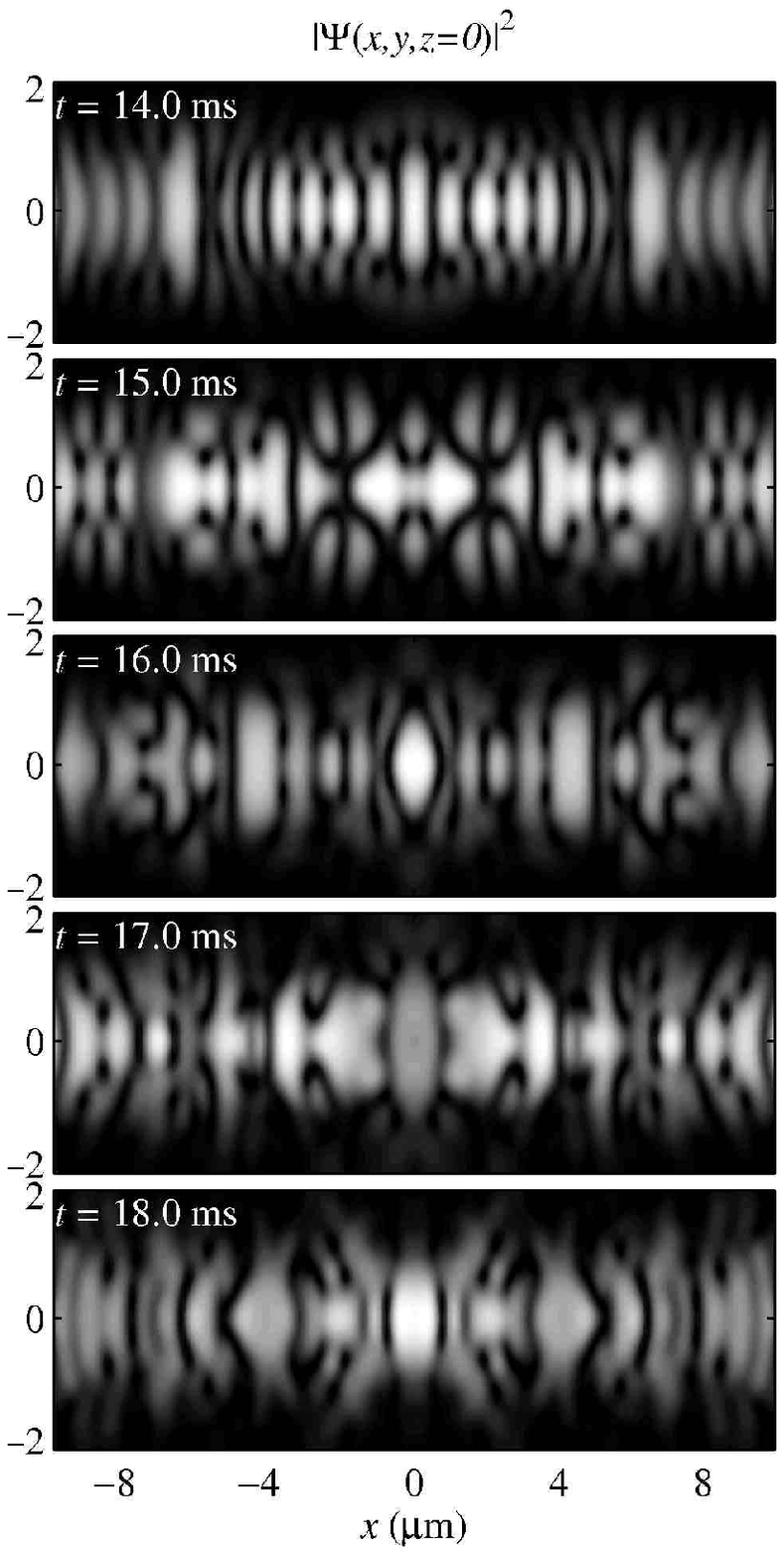}
      \center{(c)}
    \end{minipage}
  \end{tabular}
  \caption{Fully three-dimensional ($x$, $y$, $z$) numerical simulations of the
    Gross-Pitaevskii equation with the same parameters as those used
    to generate Fig.\ 1 in the main text but with $N = 10^5$ particles
    instead of $10^6$ used in the experiment.  Fig.\ (a) is the
    integrated density whereas the Figs.\ (b) and (c) represent
    density slices of the condensate by the plane $z=0$ and are zoomed
    in relative to (a).  The azimuthal symmetry of the solution is
    visually maintained over the integration times studied.  A
    transverse instability of the soliton train and vortex ring generation are clear.  No
    anti-trapped expansion was performed in the simulation.  The vertical,
    $y$-axis in the figures is measured in units of microns.}
  \label{fig:blast_hold}
\end{figure*}
\pagebreak
\begin{figure*}
  \centering
  \begin{tabular}{cc}
    \begin{minipage}{0.4725\linewidth}
      \includegraphics[width=\linewidth]{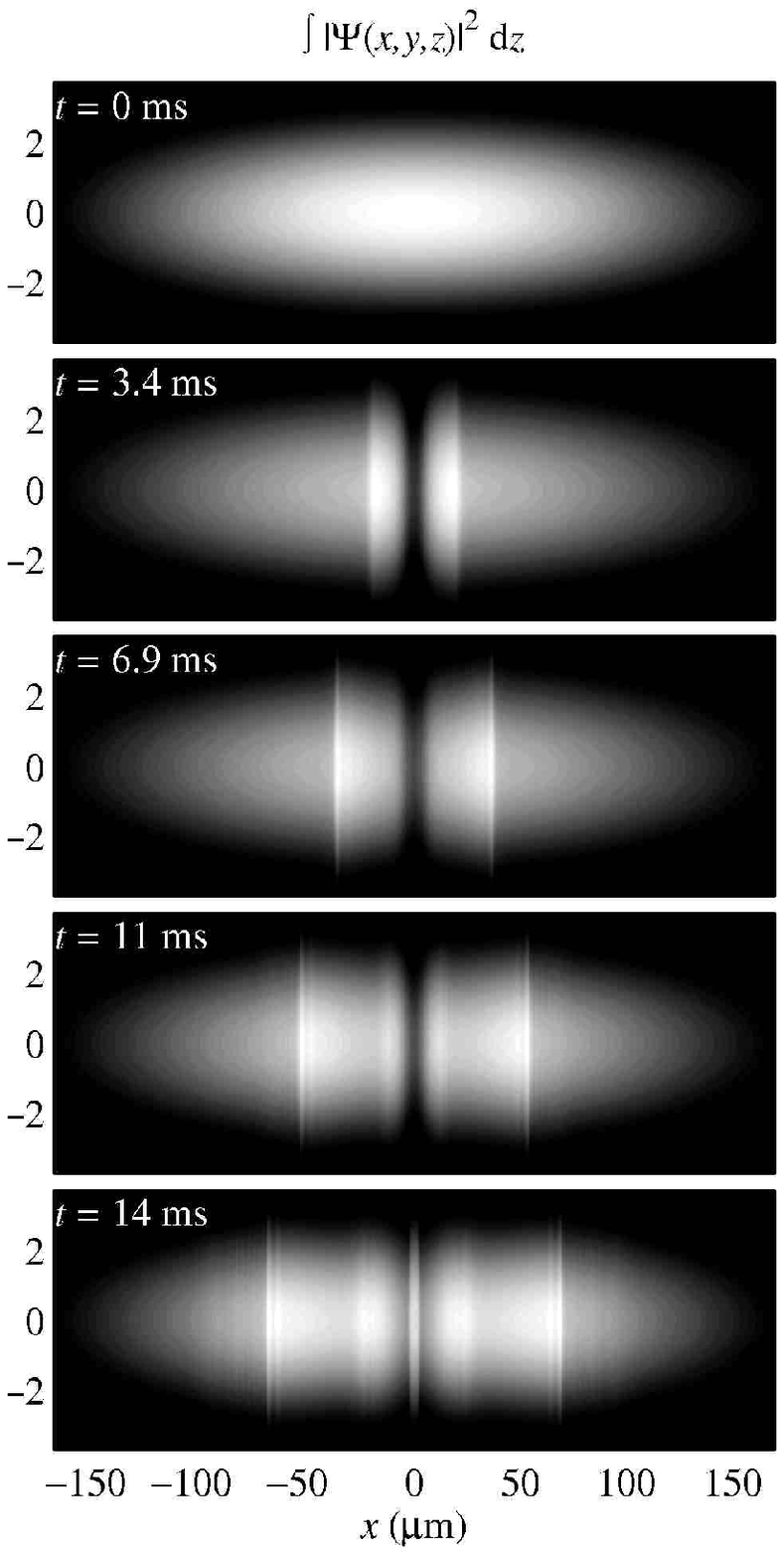}
      \center{(a)}
    \end{minipage}
    &
    \begin{minipage}{0.4725\linewidth}
     \includegraphics[width=\linewidth]{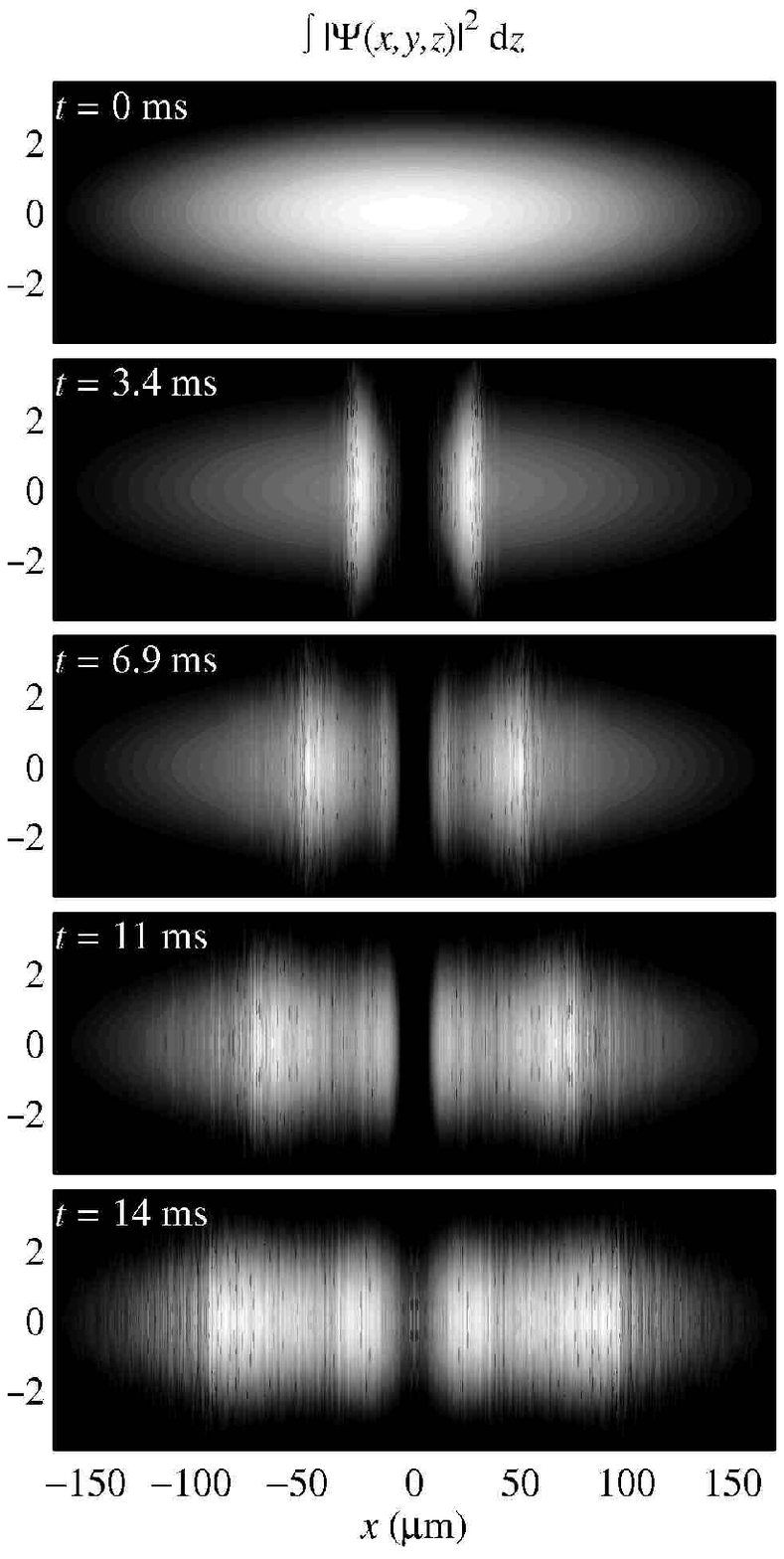}
    \center{(b)}
    \end{minipage}
  \end{tabular}
  \caption{Azimuthally symmetric numerical simulations of the
    Gross-Pitaevskii equation for the blast wave experiments.  Figs.\ (a) and (b)
    correspond to the same parameters used in the experiment shown in
    Figs.\ 2(c) and 2(d) in the main text respectively where a repulsive dipole laser was
    turned on and held on for 12 ms and then turned off for 2 ms.  The
    repulsive dipole beam for Fig.\ (a) has power $P = 360 \mu$W
    whereas the power in (b) is $P = 1.99$ mW. In the low power case (a), there is no
    transverse instability leading to vortex ring generation and therefore, the excitations are
    interpreted as sound waves.  The dark dots that emerge inside the
    condensates during the evolution in the high power case of Fig.\ (b)
    indicate vortex rings due to a transverse instability of dispersive
    shock waves. No anti-trapped expansion was performed in any simulations.  The vertical, $y$-axis in the figures
    is measured in units of microns.}
  \label{fig:blast_hold}
\end{figure*}
\pagebreak
\begin{figure*}
  \centering
  \begin{tabular}{cc}
    \begin{minipage}{0.4725\linewidth}
      \includegraphics[width=\linewidth]{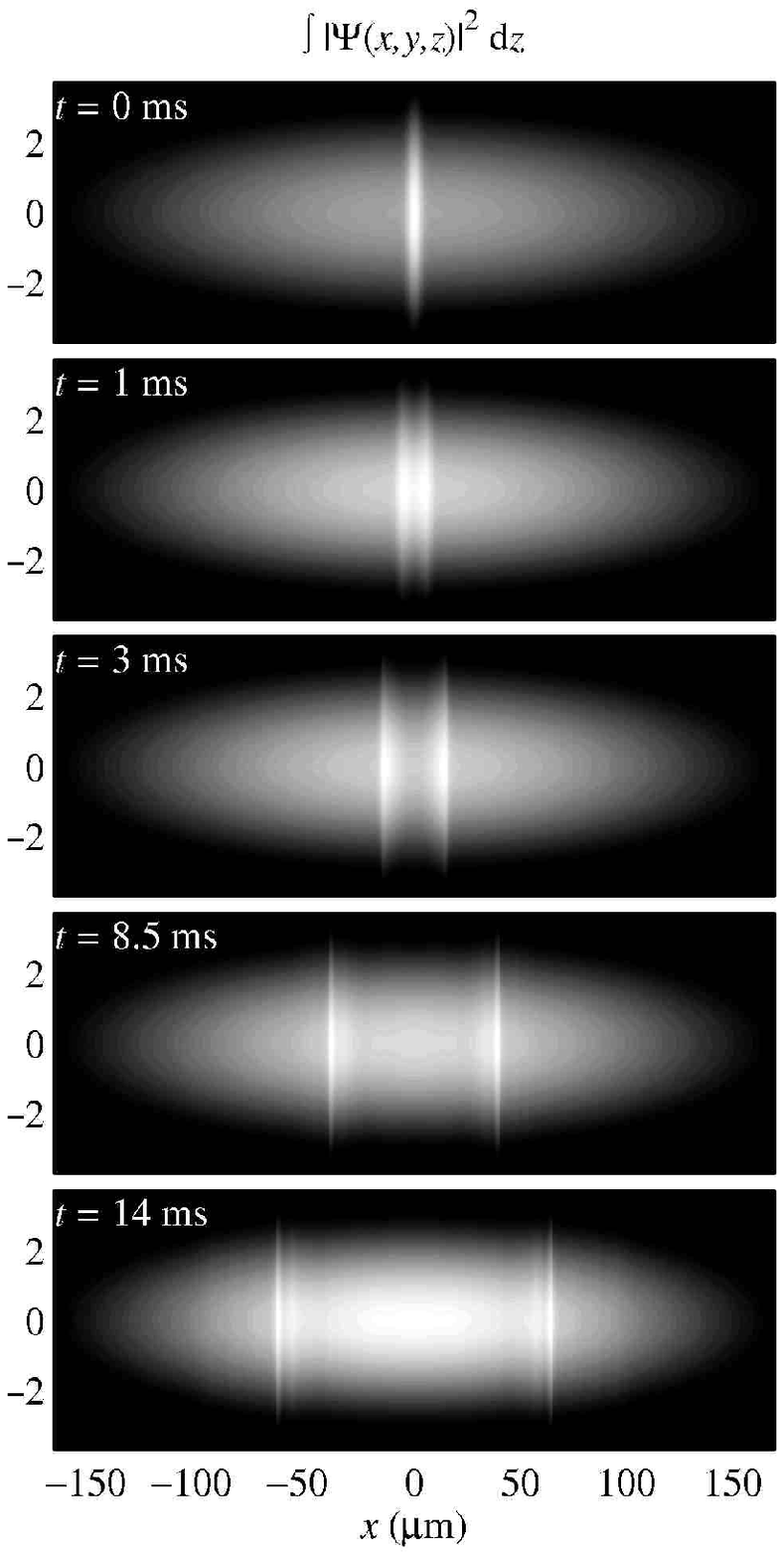}
      \center{(a)}
    \end{minipage}
    &
    \begin{minipage}{0.4725\linewidth}
    \includegraphics[width=\linewidth]{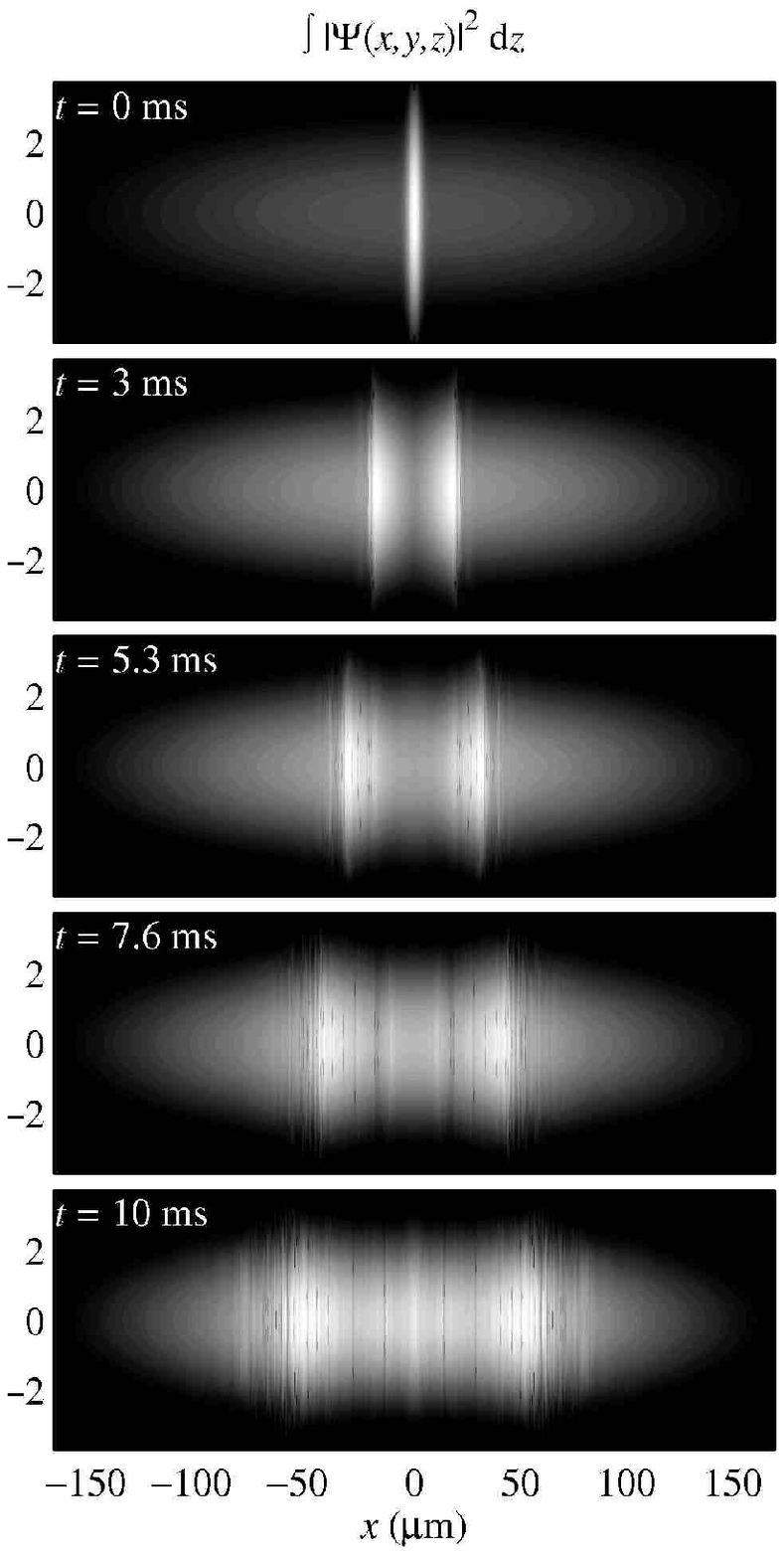}
    \center{(b)}
    \end{minipage}
  \end{tabular}
  \caption{
    Azimuthally symmetric numerical simulations of the Gross-Pitaevskii
    equation for the attractive dipole beam experiment. Fig.\ (a)
    corresponds to the same parameters used to create the experimental
    images in Figs.\ 4(a-d) in the main text (with dipole laser power $P =
    61 ~ \mu$W).  Fig.\ (b) corresponds to the same parameters used in
    the experiments shown in Figs.\ 4(e) and 4(f) (with dipole laser power $P
    = 183 ~ \mu$W). In the lower power case (a), there is no transverse
    instability leading to vortex ring generation and therefore, the
    excitations are interpreted as sound waves.  The dark dots that
    emerge inside the condensates during the evolution in the high power
    case of Fig.\ (b) indicate vortex rings due to a transverse
    instability of dispersive shock waves. No anti-trapped expansion was
    performed in any simulations. The vertical, $y$-axis in the figures is measured in
    units of microns.}
  \label{fig:attrac}
\end{figure*}


\begin{thebibliography}{}
\bibitem{footnotesoundspeed}
The two density peaks slightly slow down as they travel towards outer edges of the BEC. Therefore we fit a
polynomial of order 2 to the measurements of position vs. propagation time, which fits the data very well. We
quote the coefficient of the linear term as the central speed of sound. The peak seen in the middle of the BEC
in Fig.~\ref{picturematrix}(c) is due to merging dynamics that take place during the antitrapped expansion just
prior to imaging (the dipole barrier was off during the 2~ms long antitrapped expansion).
\end{thebibliography}
\end{document}